\documentclass[11pt]{article}
\usepackage{graphicx}
\usepackage{cs12,html,hyperref}

\begin{document}
 
\title{A Dynamical Origin for Brown Dwarfs}
 
\author{Bo Reipurth\altaffilmark{1,2}, \  Cathie Clarke\altaffilmark{3}, \ \& Eduardo Delgado-Donate\altaffilmark{3}}
\altaffiltext{1}{Center for Astrophysics and Space Astronomy, University of Colorado, Boulder, USA}
\altaffiltext{2}{Present address: Institute for Astronomy, University of Hawaii, Honolulu, USA}
\altaffiltext{3}{Institute of Astronomy, University of Cambridge, UK}

\index{brown dwarfs}
\index{multiple systems}
\index{extrasolar planets}

\begin{abstract}

Brown dwarfs may have such low masses because they are prematurely
ejected from small unstable multiple systems, while the members are
still actively building up their masses. We demonstrate that this
scenario is consistent with all currently existing observations of
brown dwarfs, and propose further observational tests. We review the
status of the latest realistic numerical simulations of disintegrating
small N clusters, which show that many of the ejected members end up
with masses that are substellar, drifting away from their birth region
with velocities rarely exceeding 2 km~s$^{-1}$.

\end{abstract}

\section{Introduction}

It is commonly assumed that brown dwarfs are formed the same way as
stars, except under conditions that lead to stellar objects with very
small masses, i.e. from clouds that are very small, very dense, and
very cold. However, with the growing realization that brown dwarfs may
be as common as stars, it is becoming disturbing that such special
physical conditions are not readily found in the molecular clouds of
our Galaxy (although they may exist elsewhere, see Elmegreen 1999).

Alternatively, Lin et al. (1998) suggested that the encounter between
two protostars with massive disks could fling out tidal filaments with
lengths of about 1000~AU, out of which a brown dwarf might form.
However, the fact that brown dwarfs are increasingly discovered also
in loose T~Tauri associations like Taurus, where such encounters
should be extremely rare, suggests that this mechanism is unlikely to
be a major source of brown dwarfs.

Taking another approach, we have recently proposed that brown dwarfs
have such extremely low masses because they were ejected from small
clusters of nascent stellar embryos (Reipurth \& Clarke 2001). This
can occur because the timescale for dynamical interactions and
ejection is comparable to the timescale for collapse and build-up of a
star. In the following, we outline the basic aspects of this model,
and summarize the status of current efforts to numerically model such
ejections.

\section{Multiplicity of Newborn Stars}

Observations over the last decade have established that young T~Tauri
stars have the same or a slightly higher binary frequency than at the
main sequence (e.g. Reipurth \& Zinnecker 1993; Ghez et al. 1993;
K\"ohler \& Leinert 1998). A small, although not very well determined,
fraction of both young and more evolved stars are also triple or
higher-order multiple systems. The situation is much less clear among
the very youngest, still embedded stars, due to the difficulties of
probing into the heavily shrouded environment of such
objects. However, new high resolution infrared techniques from the
ground and space, as well as centimeter interferometry with e.g. the
VLA, are beginning to yield results. In a detailed study of 14 driving
sources of giant Herbig-Haro (HH) flows, Reipurth (2000) found that
more than 80\% are binaries, and of these half are higher order
systems. It should be noted that these are the actually observed
frequencies, without corrections for the considerable incompleteness
of the observations, and so the results are in fact consistent with
the possibility that {\em all} giant HH flow sources may be binary or
multiple systems. These embedded outflow sources are of the order of
10$^5$~yr old or less, and it follows that some of these systems must
decay to reach the lower observed frequencies at later evolutionary
stages. It is well established that non-hierarchical triple systems
undergo rapid dynamical evolution and evolve into either a binary with
a distant companion, i.e. a hierarchical triple system, or into a
binary and an unbound, escaping third member (see Section~3). The
binary system that is formed in this dynamical process is highly
eccentric, and given that the triple disintegration is likely to take
place while the stars are still actively accreting gas from an
infalling envelope, it follows that the circumstellar disks will
interact on an orbital timescale, which will lead to shrinkage of the
orbit (e.g. Artymowicz \& Lubow 1996). These interactions are again
likely to cause cyclic variations in the accretion rate, with
consequent pulses in the outflow production, and the
giant HH flows may therefore represent a fossil record of the birth and early
evolution of binary systems (Reipurth 2000). Altogether, it appears
that the generation of giant HH flows, the birth of a binary, and the
formation of brown dwarfs may all be different aspects of a single
event, namely the dissolution of a small multiple system.

\section{Dynamical Interactions in Multiple Systems}

 If the dominant mode of star formation involves the splitting of a
core in to $N=3$ or more fragments then important new ingredients
enter the dynamics which are not encountered in the cases $N=1$ or
$N=2$. The reason for this is simply that most $N>2$ body
configurations are dynamically unstable; over several dynamical
timescales the components exchange energy through the action of
gravitational forces until they attain a stable equilibrium. In
practice, this generally means that the mini-cluster disintegrates,
the energy for this process being derived from the formation of at
least one binary system. After a number of dynamical timescales,
therefore, the core no longer consists of a bound stellar system, but
of a mixture of binary and single stars that drift apart so as to
mingle with the ambient stellar field.

\begin{figure}[t]
\hspace{0.9cm}
\hspace{0cm}
\vbox{\includegraphics*[angle=270, width=11.0cm]{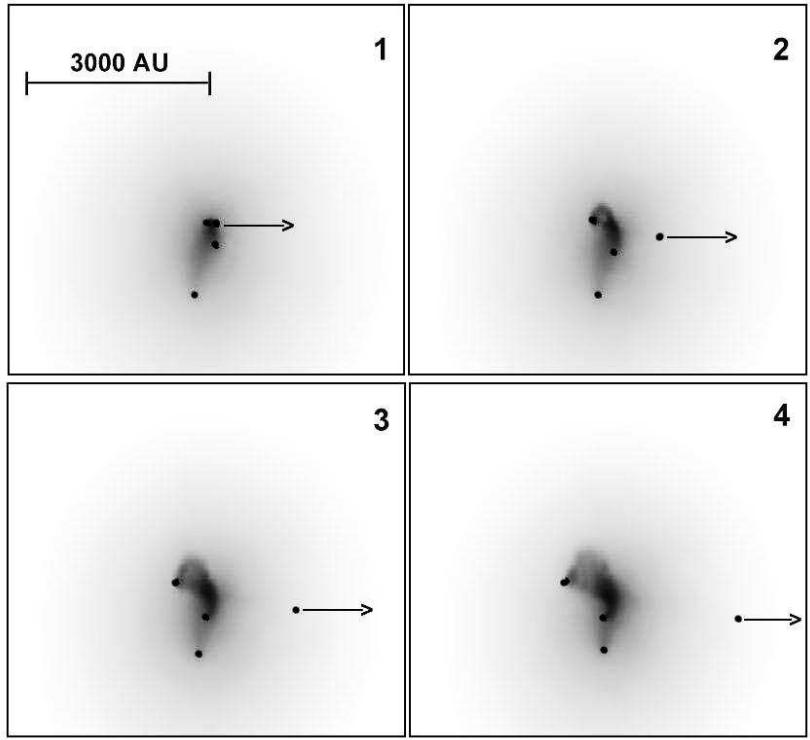}
\vspace{-0.5cm}}
\caption{\footnotesize{
Snapshots showing the ejection of a $0.06M_\odot$ brown dwarf with a
final speed of $\sim 1.5~kms^{-1}$, from a simulation by
Delgado-Donate. The sequence, which covers a time span of 
$\sim 10,000~yr$, displays the typical features of this kind of models: 
the escape of a light object expelled from an unstable multiple system 
together with the subsequent recoil of its tightest bound members. The brown
dwarf has been highlighted with an arrow.}}
\end{figure}

  Below we summarise the progress that has been made to date in
quantifying such behaviour. We would like to know what is the mass
spectrum of the stars (and brown dwarfs) that are produced in this
way, what are their binary statistics, how much circumstellar material
young stars can bring with them as they emerge from such a core? We
would also like to know what are the kinematic properties of the
resulting stars and brown dwarfs. Unfortunately, the answers to these
questions can be most readily obtained in the case of simulations that
omit a vital ingredient (i.e. gas); and it is only recently that it
has been possible to address these issues in hydrodynamical
simulations. Even so, the twin demands of simulations that are both
well resolved and can be performed a large number of times (in order
to obtain statistically meaningful results) takes us close to the
limits of what is computationally viable at the present time. Here we
describe the results of simulations that are progressively more
realistic, starting with pure N-body experiments and ending with what
represents the current state of the art - gas dynamical simulations
starting from turbulent initial conditions.  Although these results
will give a flavour of the physical processes involved, it should be
stressed that we are not yet at a stage where statistical results can
be derived from the most realistic experiments.  This inevitably means
that the observational predictions (described in Section 4) remain
somewhat provisional.

      The simplest possibility conceptually (which is however not
likely to be even approximately true in practice) is if all the gas is
instantly accreted on to each protostar so that the ensemble evolves
thereafter as a system of point masses. This situation is
straightforward to model as an Nbody system and has been analysed by
many authors (e.g. van Albada 1968; Sterzik \& Durisen 1998). The usual
outcome of such simulations is that the two most massive members of
the system form a binary, whereas the remainder are ejected as single
stars. The ejection velocity of stars is related to their orbital
velocity during a close three body encounter. For example, in Sterzik
and Durisen's simulations, the typical separation between stars is
initially around $100$ AU, and typical ejection velocities of single
stars are 3-4 km s$^{-1}$. If the same system of point masses had been
set up with separations a factor $10$ greater (say), the resulting
ejection velocities would be a factor ${10}^{1/2}$ smaller. Sterzik
and Durisen also quantified the dependence of ejection velocity on
stellar mass, and found that for $N>3$, such a dependence is very
weak; they however found a significant difference in the final
velocities of single stars and binaries, with the centre of mass
velocity of binaries being typically a factor $3-6$ less than that of
singles.

   The above simulations have all the advantages of computational
simplicity and yield well defined predictions for the fraction of
stars of various masses that end up in binaries (McDonald \& Clarke
1993).  They however take no account of the fact that the fragments
will not interact as point masses but will instead have their
interactions mediated by the presence of circumstellar disks. This
situation is less straightforward to model numerically, since there
are well known numerical difficulties in maintaining disks for many
internal orbital timescales against the dispersive action of
viscosity.  Some insight into the expected role of disks can be
obtained from the calculations of McDonald \& Clarke (1995), who
included the effect of star-disk interactions in a parameterised form
(see also Clarke \& Pringle 1991). The main role of disks is to
harden temporary binaries so as to protect them against disruptive
encounters with other cluster members. As a result, more than one
binary can be formed in each cluster; although the most massive
cluster member is always in a binary, its companion is now picked at
random from the cluster members. Thus the net effect of star-disk
interactions is to boost the numbers of lower mass stars that end up
in binaries (either as primaries or secondaries) relative to the
dissipationless (Nbody) case.

   However, neither of the sorts of simulations described above can
say anything about the stellar initial mass function, since stellar
masses are assigned at the outset of the simulation. Such
instantaneous mass assignment is of course a very poor approximation
to the behaviour of real fragmenting cores: the interactions that lead
to the formation of binaries and break up of the cluster occur over a
few dynamical times, but the infall of mass onto each of the stars
happens on a comparable timescale. Thus realistic simulations need to
address the whole process as a hydrodynamic one and follow through the
evolution of a core that is initially $100 \%$ gas.
 
  To date, some progress has been made on this issue by setting up
cores in which `seeds' of collapsed gas have been planted (Bonnell et
al 1997; Bonnell et al. 2001). These seeds grow in mass due to gas
accretion in an inequitable manner - their orbital histories determine
whether they spend much time in the densest central regions of the
core and hence how much mass they acquire.  Such simulations vividly
demonstrate how `competitive accretion' works: seeds that get an early
headstart in the race for mass tend to settle into the cluster core
and thereby acquire more mass, whereas seeds that do not grow much
initially are more likely to be flung out of the core and hence be
prevented from further growth.  Thus competitive accretion provides a
ready mechanism for obtaining a large dynamic range of final stellar
masses from arbitrary initial conditions.

\begin{figure}[t]
\hspace{-0.6cm}
\hspace{0cm}
\vbox{\includegraphics*[angle=90, width=13.0cm]{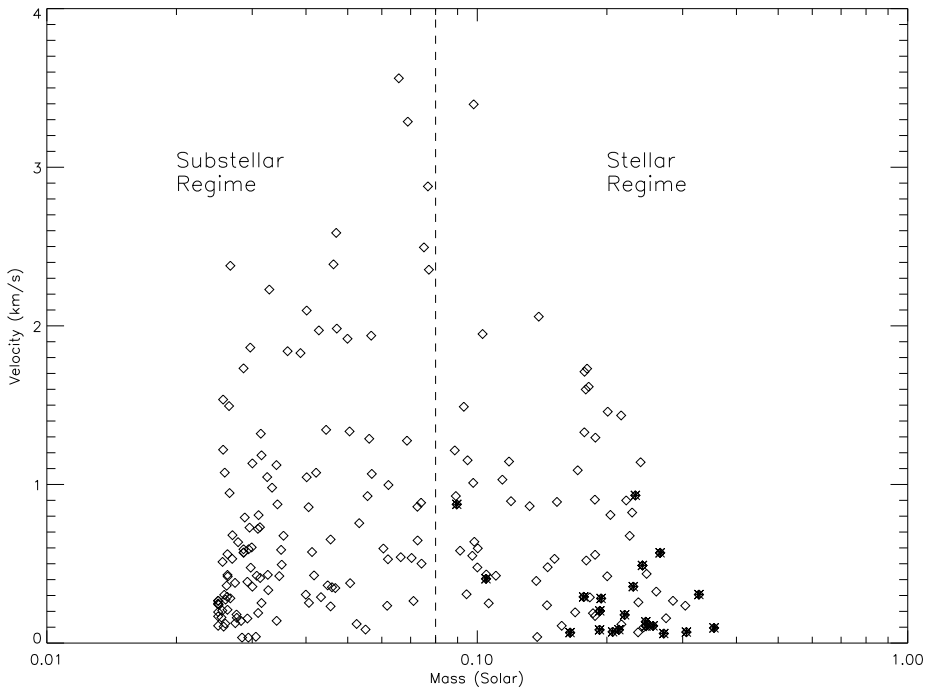}
\vspace{-0.3cm}}
\caption{\footnotesize{The final velocity of ejected stars (diamonds = singles,
asterisks= binaries) versus mass, from the simulations of Delgado-Donate.
These simulations model cores of mass $0.5 M_\odot$ consisting
initially of 5 `seeds' each of mass $0.025 M_\odot$, and the
remaining $75 \%$ of the core mass in gas. The initial virial
velocity of the cores is $\sim 0.2$ km~s$^{-1}$.} }
\end{figure}

  Recent simulations by Delgado-Donate (in preparation) have begun to
quantify the IMF produced by competitive accretion during the break up
of small clusters (Figure~1) and find that it is broadly compatible
with the observed IMF.  (Note that in these simulations no disks are
formed around the protostars due to the absence of small scale
turbulence in the initial core. Thus although these simulations model
the accretion of gas on to the stars as they form, and the consequent
shrinkage of the protostellar separations, they do not model the
effects of star-disk interactions).  The dynamical outcome of such
disintegrations shares many qualitative similarities with the
dissipationless (Nbody) results of Sterzik and Durisen. As in their
simulations, there is no appreciable dependence of final velocity on
resulting stellar mass (see Figure 2), but the binaries attain speeds
that are a factor $\sim 5$ less than the typical ejection speeds of
the single stars. One notable difference, however, is the relationship
between the typical ejection speeds and the initial parameters of the
core. In the dissipationless simulations, the final velocities of
ejected stars are of the same order as the virial velocity of the
initial core. In the gas dynamical simulations, there are also many
objects that are ejected early on with such velocities, as shown in
Figure~2, where the initial virial velocity of the parent core is
around $0.2$ km~s$^{-1}$. However, Figure~2 also shows that many stars
are ejected with velocities much greater than this: about half the
stars attain velocities greater than $3$ times the initial virial
velocity (the corresponding number for Sterzik and Durisen's
simulations is less than $1 \%$) and there is also a significant
minority that attain velocities more than a factor $10$ greater than
the initial virial value.  This may be readily understood inasmuch as
in the gas dynamical simulations, the systems shrink as the gas is
accreted, so that later interactions are much closer and produce
correspondingly larger velocity kicks.  Thus in these gas dynamical
simulations (where the typical initial separations of protostars are
$10^4$ AU) , the ejection velocities are similar to Sterzik and
Durisen's dissipationless calculations in which the stars are
initially separated by only of order $100$ AU.

  These simulations also give some insight into the binary pairing
characteristics of the resulting stars. The outcome of each five star
simulation is the production of a binary (sometimes a triple). (Note
that these simulations do not give reliable results for the
separations of the resultant binaries, since they do not include
star-disk interactions, which will tend to harden the binaries
considerably).  It is notable that these simulations produce extreme
mass ratio pairs (e.g. $10:1$) quite readily, which contrasts strongly
with pure Nbody results, where such an outcome is almost unknown.

  A further step towards reality can be obtained by abandoning the
artificial distinction between seeds and smoothly distributed
background gas in the above simulations. A more physical approach is
to start with gas that is subject to a supersonic turbulent velocity
field which rapidly generates a richly non-linear density structure in
the gas (Klessen et al. 2000; Klessen 2001).  Such simulations follow
not only the competitive accretion between contending `stars' and
their dynamical interactions, but also the very formation of the stars
from pockets of Jeans unstable gas. The most ambitious simulation to
date is that of Bate (see Bate et al. 2001) which models a system that
will form of the order of a hundred stars, with the capacity to resolve
structures forming down to mass limits of a few Jupiter masses. This
system readily demonstrates the formation of small N ensembles in
which the sort of behaviour described above (binary formation,
competitive accretion, ejection of low mass members) is observed to
occur. Such a one-off simulation cannot be used to generate
statistical results however, and so follow up simulations, tracing the
evolution of large numbers of small N ensembles are required.

\section{Observational Tests}

{\em Brown Dwarfs and Early Stellar Evolution}

Small N-body systems still in the process of accumulating mass from an
infalling envelope would observationally be seen as a Class~0 or
perhaps a Class~I source with strong outflow activity.  If brown
dwarfs are formed by the disintegration of small N-body systems, it
follows that the very youngest, indeed newborn, brown dwarfs should be
found in the immediate vicinity of such sources. As a small triple or
multiple system breaks up, its members start to drift out of the
nascent envelope, and may on relatively short time scales (of order a
couple of thousand years) emerge from being deeply embedded infrared
sources with ample far-infrared and sub-mm emission to being optically
visible T Tauri like stars. In this radically different picture of
early stellar evolution, the gradual and smooth transition between
Class~0 and Class~II sources can be replaced by a rather abrupt
transition, and the main accretion phase for the members of a multiple
system is terminated not by the infalling envelope running out of gas,
or outflow blowing away the last parts of the envelope, but by the
newborn members ``leaving the nest'' (Reipurth 2000). Considering the
rapidly increasing evidence for multiplicity among the youngest stars,
it is interesting to review the early idea by Larson (1972) that all
single low mass stars may have formed in dynamical interactions among
newly born multiple systems.

One observational test of the dynamical formation model of brown
dwarfs would be to study carefully the statistics of brown dwarfs in
the vicinity of Class~0 sources.  To get a practical sense of the
observed separation between a nascent brown dwarf and its siblings,
assume that it is observed at a time $t$ [yr] after ejection and
moving with the space velocity $v$ [km s$^{-1}$] at an angle $\alpha$
to the line-of-sight in a star forming region at a distance $d$
[pc]. Then the projected separation $s$ in arcsec is $ s =
0.21~v~t~d^{-1}~sin\alpha. $ For a velocity of 1 km~s$^{-1}$, a brown
dwarf moving out of a nearby ($d$ $\sim$ 130 pc) cloud at an angle of
60$^o$ to the line-of-sight will already be 5 arcmin away after 2
$\times$ 10$^5$ yr.  Also, half of all ejected brown dwarfs will move
into the cloud from which they formed. Such objects will be detectable
only as highly extincted and weak infrared sources.

\vspace{0.1cm}
\noindent
{\em Brown Dwarfs with Stellar Companions}

It has been known for some time that brown dwarfs are only rarely
found as close (less than 3~AU) companions to low mass stars (the
``brown dwarf desert''). But recent work by Gizis et al. (2001) has
demonstrated that brown dwarfs are commonly companions to normal stars
at large separations (greater than 1000~AU). The separation
distribution function of brown dwarfs in binary systems contains
important information about their formation, and establishing its form
more precisely will form a crucial test for any theory of brown dwarf
formation. The ejection hypothesis readily explains the currently
available observations: brown dwarfs should rarely be found as close
companions to stars, as in this environment they would, except for
special circumstances, have continued to accrete mass at almost the
same rate as their stellar companions, thus pushing through the
substellar/stellar boundary at almost the same time. On the other
hand, distant brown dwarf companions are readily expected, because not
all ejections will lead to unbound systems.

A further test of the ejection scenario comes from studies of the
binarity of brown dwarfs. A number of brown dwarfs have been found to
be binaries (e.g. Mart\'\i n, Brandner, \& Basri 1999), but,
intriguingly, they appear to be rather close binaries, whereas wide
pairs (many hundreds of AU) have so far not been found. In the
ejection scenario, if two stellar embryos are ejected as a pair, they
have to be relatively close in order to survive as a binary. Obviously,
the precise limit for survival depends not only on the mass of the
pair, but also on whether the ejection occurs from a compact or a
wider configuration of embryos. Numerical simulations will help to
quantify this.

\vspace{0.1cm}
\noindent
{\em Kinematics of Brown Dwarfs}

When a small N-body cluster dissolves, its members drift apart and
blend with the other young stars in the general star forming
environment, whether it is an association or a rich cluster. The
velocities of young stars are generally assumed to be similar to the
turbulent velocity of the gas out of which they formed, i.e. of the
order of a km per second. If all stars are formed in small multiple
systems (Larson 1972), then to this we must add the mean velocity of
the dispersing members. 

If brown dwarfs are ejected stellar embryos, they must carry kinematic
evidence reflecting their origin in small multiple systems. To first
order, the ejected member from such a system acquires a velocity
comparable to the velocity attained at pericenter in the close triple
encounter. An observational test of the ejection scenario would
therefore be to compare observed velocities of brown dwarfs with
velocities derived in realistic numerical simulations.  However, the
brown dwarfs studied must be young and belong to loose
associations. In denser clusters, two body relaxation will soon
dominate the kinematics of its members, and field brown dwarfs will be
dominated by objects that have evaporated from clusters and thus
are similarly affected.

As discussed in Section~3, and illustrated in Figure~2, the most
recent, realistic simulations show that the majority of brown dwarfs
are ejected with space velocities of less than 1 km~s$^{-1}$, and only
a small fraction have velocities larger than
2~km~s$^{-1}$. One-dimensional velocities (radial or tangential) are
correspondingly smaller. 

It follows that a kinematical test of the ejection scenario will be
very difficult to carry out. In the first kinematical study of brown
dwarfs, Joergens \& Guenther (2001) found that the {\em radial}
velocity dispersion of nine brown dwarfs in Cha~I is 2.0~km~s$^{-1}$,
whereas for T~Tauri stars in the same region it is
3.6~km~s$^{-1}$. These numbers should be compared to the velocity
dispersion of the gas, which is 1.2~km~s$^{-1}$ in the region. As
illustrated in Figure~2, there is virtually no dependence of the
ejection velocity on mass, i.e. if most or all single stars have been
ejected, then there should be little if any kinematic difference
between brown dwarfs and surrounding young stars. The existing,
limited observations are thus consistent with our current
understanding of the kinematics of ejected brown dwarfs and low mass
stars. 

One possible kinematic signature which may be unique to the ejection
scenario is that the binaries that recoil from a dissolving small N
cluster will have significantly smaller velocities than the ejected
single objects. Once a young stellar association has been carefully
studied to identify all binaries, this could be an interesting test.

\vspace{0.1cm}
\noindent
{\em Signatures of Youth in Brown Dwarfs}

The expulsion of an ``unfinished'' hydrostatic core from an infalling
envelope may place limits on the amount of circumstellar material that
can be brought with it. Freefloating brown dwarfs will therefore have
{\em finite} reservoirs from which they can accrete and form
planets. However, it is only the outermost regions of a circumstellar
disk that will be truncated, the precise values depending on the impact
parameters in the close triple encounter that triggers the
expulsion. The innermost disk regions will be unaffected, and it is
these regions that are responsible for the near-infrared excesses and
T~Tauri like characteristics that are observed in very young brown
dwarfs (e.g. Muench et al. 2001). Brown dwarfs are therefore expected
to show mostly the same characteristics of youth as T~Tauri stars, and
may even be able to form planets, but because of the limited amount of
gas they carry with them, the period during which they display T~Tauri
like characteristics may be more short-lived than for their heavier
stellar siblings.

\section{Brown Dwarfs and Extrasolar Planets}

 The discovery in recent years of numerous giant extrasolar planets
orbiting other stars (e.g. Marcy, Cochran \& Mayor 2000), as well as a
large number of low-luminosity objects that have been suggested to be
freefloating giant planets (e.g. Lucas et al. 2001) suggests that a
continuum may exist between such giant planets and brown dwarfs. By
its definition as an object that cannot attain hydrogen burning, a
brown dwarf must have a mass of less than about 0.08 M$_\odot$, or
80~M$_{Jupiter}$, weakly dependent on metallicity (Chabrier \& Baraffe
2000). Most of the extrasolar planets found around other stars have
masses of order {\em M}sin{\em i} $\leq$ 10~M$_{Jupiter}$. The
difference between the masses of giant extrasolar planets and brown
dwarfs is therefore about an order of magnitude, much less than the
three orders of magnitude spanned by stellar masses. With no clear
dividing line between brown dwarfs and giant extrasolar planets, it is
thus of considerable interest to ask which, if any, differences exist
between these two classes of objects, and whether they perhaps could
have a common origin.

It is indeed a possibility that freefloating giant planets are formed
the same way as brown dwarfs, i.e. as ejectae from unstable multiple
systems of forming stars. Boss (2001) has advocated precisely such a
scenario, suggesting that magnetic field tension in the collapse
process has the effect of splitting the infalling material into very
closely spaced fragments, from which objects as small as a
Jupiter-mass can be immediately ejected by dynamical interactions.
Another possibility is that freefloating giant planets are formed like
the giant planets in our own solar system, i.e. by later condensing
out of material in a circumstellar disk. Subsequently, dynamical
relaxation may ultimately kick one or more of the giant planets out of
their nascent planetary system (e.g. Papaloizou \& Terquem
2001). Obviously both mechanisms, despite their very different nature,
may create freefloating planetary sized objects, which observationally
do not appear to be easily distinguishable.  Because ejected objects
lose most traces of their pre-history by their expulsion, attempts to
compare brown dwarfs and giant extrasolar planets should not be done
on the basis of freefloating objects.

On the other hand, bound systems, in which giant planets or brown
dwarfs are orbiting a star, may retain critically important memories
of their formation processes. And for such objects major differences
exist between the two classes of objects. As discussed earlier, brown
dwarfs are commonly found bound to stars only at very large
separations. Their formation history clearly do not commonly permit
them to exist closely bound to a star, as expected and readily
explained by the ejection scenario.  This is in marked contrast to the
increasingly well determined statistics for giant extrasolar planets
orbiting other stars, which show that such objects are often located
at extremely small separations, in regions that are off-limits to
brown dwarfs in the ejection scenario. If the observed separations
were a smooth function of mass between the two categories of objects,
one could possibly have argued that orbital evolution had led the
smallest mass objects to migrate to locations close to their
stars. But given that the two populations are strictly separated in
the way they orbit their stars, we conclude that brown dwarfs and
giant extrasolar planets, although both arguably having sizable
freefloating populations, are likely to be formed in very different
manners. In this interpretation, giant planets are formed at later
times out of circumstellar disks and their growth is limited by how
much disk material is available, whereas brown dwarfs are failed
stars, which due to their expulsion could not take full advantage of
their rich supply of inflowing gas to grow beyond the hydrogen burning
limit.

\acknowledgments

BR thanks Observatoire de Bordeaux, where part of this review was
written, for hospitality.


\vspace{-0.3cm}



\begin{references}

\vspace{-0.2cm}

\footnotesize{

\reference Artymowicz, P., Lubow, S.H. 1996, ApJ, 467, L77

\reference Bate, M.R, Bonnell, I., \& Bromm, V. 2001, {\em The Origins of Stars and Planets: The VLT View' ESO Astrophysics Symposia}, Eds. J. Alves, M.J. McCaughrean (Springer), in press

\reference Bonnell, I.A., Bate, M.R., Clarke, C.J. \& Pringle, J.E.
 1997, MNRAS 323,785

\reference Boss, A.P. 2001, ApJ, 551, L167 

\reference Chabrier, G., Baraffe, I. 2000, ARAA, 38, 337

\reference Clarke, C. \& Pringle, J.E. 1991, MNRAS 249, 588

\reference Elmegreen, B.C. 1999, ApJ, 522, 915

\reference Ghez, A.M., Neugebauer, G., Matthews, K. 1993, AJ, 106, 2005

\reference Gizis, J.E., Kirkpatrick, J.D., Burgasser, A., Reid, I.N., Monet, D.G., Liebert, J., Wilson, J.C. 2001, ApJ, 551, L163

\reference Joergens, V., Guenther, E. 2001, A\&A, in press

\reference Klessen, R. 2001, ApJ 556, 837

\reference Klessen, R., Heitsch, F., Mac Low, M.  2000, ApJ 535, 887

\reference K\"ohler, R., Leinert, C. 1998, A\&A, 331, 977

\reference Larson, R.B. 1972, MNRAS, 156, 437

\reference Lin, D.N.C., Laughlin, G., Bodenheimer, P., Rozyczka,
M. 1998, Science, 281, 2025

\reference Lucas, P.W., Roche, P.F., Allard, F., Hauschildt, P.H. 2001,
MNRAS, 326, 695

\reference Marcy, G.W., Cochran, W.D., Mayor, M. 2000, {\em Protostars
and Planets IV}, Eds. V. Mannings, A.P. Boss, S.S. Russell, (Univ. of
Arizona Press), p.1285

\reference Mart\'\i n, E.L., Brandner, W., Basri, G. 1999, Science, 283, 1718

\reference McDonald, J. \& Clarke, C.J. 1993, MNRAS 262, 800

\reference McDonald, J. \& Clarke, C.J. 1995, MNRAS 275, 671 

\reference Muench, A.A, Alves, J.A., Lada, C.J., Lada, E.A. 2001, ApJ, 558, L51

\reference Papaloizou, J.C.B. \& Terquem, C. 2001, MNRAS, 325, 221

\reference Reipurth, B. 2000, AJ, 120, 3177

\reference Reipurth, B., Zinnecker, H. 1993, A\&A, 331, 977

\reference Reipurth, B. \& Clarke, C.J. 2001, AJ, 122, 432

\reference Sterzik, M. \& Durisen, R. 1998, A\&A 339, 95

\reference van Albada, T.S. 1968, Bull. Astron. Inst. Netherlands, 19, 479

\reference Woitas, J., Leinert, Ch., K\"ohler, R. 2001, A\&A, 376, 982
   
}

\end{references}
\end{document}